\documentclass[%
 aip,
 rsi,
amsmath,amssymb,
nobibnotes, 
reprint,
longbibliography
]{revtex4-1}

\usepackage{threeparttable}
\usepackage{graphicx}
\usepackage{dcolumn}
\usepackage{bm}

\usepackage[utf8]{inputenc}
\usepackage{mathptmx}
\usepackage{multirow}
\usepackage{braket}
\usepackage{tablefootnote}

\usepackage[english]{babel}

\usepackage{xcolor}
\colorlet{RED}{red}
\colorlet{BLUE}{blue}
\usepackage{dcolumn}
\usepackage{bm}
\usepackage[version=4]{mhchem} 
\usepackage{acronym}
\usepackage{adjustbox}
\usepackage{float}
\usepackage{tikz}
\usepackage{microtype} 
\usepackage{algorithm}
\usepackage{enumitem}
\usepackage{appendix}
\usepackage{amsmath}
\usepackage{algpseudocode}
\usepackage[caption=false]{subfig}

\usetikzlibrary{calc,shapes.geometric,decorations.pathmorphing,patterns}

\definecolor{background-color}{gray}{0.98}
\usepackage[margin=2.3cm,bmargin=1cm,footnotesep=1cm]{geometry}

\usepackage{hyperref}
\hypersetup{
  colorlinks=true,
  linkcolor=blue,
  citecolor=blue,
  urlcolor=blue,
}

\begin{document}

\title{Quantum Flow algorithm: quantum simulations of chemical systems using reduced quantum resources and constant depth quantum circuits
}

\author{Bhumika Jayee}
\affiliation{%
  Physical Sciences Division,
  Pacific Northwest National Laboratory, Richland, Washington, 99354, USA
}

\author{Nathan M. Myers}
\affiliation{%
Advanced Computing, Mathematics, and Data Division,
  Pacific Northwest National Laboratory, Richland, Washington, 99354, USA
}

\author{Duo Song}
\affiliation{%
  Physical Sciences Division,
  Pacific Northwest National Laboratory, Richland, Washington, 99354, USA
}

\author{Eric J. Bylaska}
\affiliation{%
  Physical Sciences Division,
  Pacific Northwest National Laboratory, Richland, Washington, 99354, USA
}

\author{Karol Kowalski}
\affiliation{%
  Physical Sciences Division,
  Pacific Northwest National Laboratory, Richland, Washington, 99354, USA
}
\affiliation{%
  Department of Physics,
  University of Washington, Seattle, Washington,  98195, USA
}
\author{Nicholas P. Bauman}
\email{nicholas.bauman@pnnl.gov}
\affiliation{%
  Physical Sciences Division,
  Pacific Northwest National Laboratory, Richland, Washington, 99354, USA
 }

\date{July 2025}

\maketitle

\noindent
\textbf{Abstract} 
We assess the performance of the Quantum Flow (QFlow) algorithm employing cost-effective solvers based on the unitary coupled-cluster ansatz with single and double excitations (QFlow-SD). The resulting energies are benchmarked against those obtained with an analogous QFlow formulation defined in the same active spaces but augmented by higher-rank excitations, including triples and quadruples (QFlow-SDTQ). Across all molecular systems considered, QFlow-SD exhibits close agreement with results from the canonical unitary coupled cluster with singles and doubles framework, while requiring substantially fewer qubits than the latter.
For the water molecule in the cc-pVTZ basis, we further demonstrate the performance of a composite two-step downfolding strategy. In this approach, an initial coupled-cluster downfolding based on the double unitary coupled-cluster ansatz is followed by a QFlow treatment within the resulting target space, illustrating the effectiveness of combining classical downfolding with quantum flow optimization.

\section{Introduction}
The ongoing progress in quantum hardware development, error mitigation software, and error correction algorithms opens the potential for electronic structure applications on quantum computers that involve large numbers of correlated electrons. 
The shift in capabilities looming on the horizon, from experiments involving 25–50 qubits to those with 100 qubits or more, raises an important question about the most productive utilization of these systems from the standpoint of calculated energy accuracy.
\cite{babbush,cerezo2021variational,mccaskey2019quantum,100qubits,berezutskii2025tensor,goings2025molecular,barison2025quantum,kaliakin2025accurate} For direct-type approaches utilizing bare Hamiltonians in active spaces whose sizes are commensurate with the number of logical qubits, this shift may not be sufficient for achieving chemical accuracies in describing energies of correlated systems. The main reason for this situation is a dichotomy in quantum resource size required to provide a balanced description of both static and dynamical correlation effects. While static correlation effects are the main target of early quantum computing applications in chemistry, their dynamical counterpart requires a substantially larger number of qubits and cannot be properly described using current quantum computers. For systems dominated by the dynamical correlation effects, the direct approach, when active space bare Hamiltonians are used to evaluate the energy, the quantum simulations can recover only a small fraction of the correlation energy for larger systems. This raises the question of how to address these issues efficiently utilizing existing quantum computers to deliver high-accuracy results.  

The recent progress achieved in the development of various coupled cluster (CC) downfolding formalisms 
\cite{kowalski2018properties,huang2023leveraging,shee2024static}
has the potential to put quantum simulations in a different context compared to traditional approaches to quantum simulations based on the utilization of a bare Hamiltonian in a given basis set. For example, CC downfolding schemes were shown to be able to capture dynamical correlation effects through the effective Hamiltonians acting in small-sized active spaces amenable to quantum hardware simulations utilizing available quantum hardware. As recently demonstrated \cite{bauman2026coupledx}, these formulations can pave the way for quantum simulations of larger systems in realistic basis sets. 

A factor determining the efficiency of the CC downfolding procedure is the ability to accurately approximate the external cluster operator ($\sigma_{\rm ext}$) correlating active space with the remaining part of the corresponding active space. The rule of thumb is that the smaller the active space, the more information needs to be encoded in $\sigma_{\rm ext}$. In other words, the CC downfolding requires some of the pre-existing knowledge about the sought-after electronic state. The quantum flow (QFlow) procedure \cite{kowalski2023quantum} has been introduced to alleviate these problems. QFlow integrates a number of active space problems (in the vein of the so-called Equivalence Theorem \cite{kowalski2021dimensionality}) with the ideas of CC downfolding, where external cluster operators for an active-space problem are defined by cluster amplitudes originating in other active-space problems and not belonging to the active space of interest. This approach has been demonstrated to be effective even in situations when amplitudes involved in the flow are not a priori known.

Furthermore, in analogy to the multi-step renormalization procedures we have considered a composite where (1) CC downfolding is used to create effective Hamiltonians in active spaces that are still too big for direct utilization of quantum algorithms (target active space (TAS)) but $\sigma_{\rm ext}$ can be effectively approximated, (2) the QFlow procedure is used as a solver for large-size active space, where the problem is re-represented in the form of small-size active spaces belonging to TAS. 

In this Communication, we will explore several questions regarding the effectiveness of QFlow procedures where the VQE solvers for each active space included in the flow are defined by low-rank excitations readily mapped to the quantum computers. We also compare the quality of the QFlow results with the energies of the VQE solvers employing the unitary coupled cluster (UCC) parametrization and the same manifold of excitations in the full target space. It is worth noting that in these experiments, the qubit requirement of VQE-UCC simulations is significantly larger than that for the QFlow simulations. Additionally, we demonstrated the performance of the composite workflows of the previous paragraph in applications to chemical systems in realistic basis sets of the cc-pVTZ quality \cite{dunning1989gaussian}. 

\section{Theory}
While the CC downfolding and QFlow formalisms have previously been discussed in the literature, for completeness, we provide a review of their basic tenets here.

\subsection{CC downfolding}
The CC downfolding is an extension of the sub-system embedding sub-algebras (SES) theorem \cite{kowalski2018properties} valid for the single-reference CC Ansatz stating that the CC energies can be obtained by diagonalizing non-Hermitian effective Hamiltonians defined for certain types of active spaces. Its unitary CC extension, used to construct Hermitian form of effective Hamiltonians, uses the so-called double unitary coupled cluster (DUCC) Ansatz for the ground-state wavefunction $|\Psi\rangle$,
\begin{equation}
|\Psi\rangle = e^{\sigma_{\rm ext}} e^{\sigma_{\rm int}} |\Phi\rangle \;,
\label{eq1}
\end{equation}
where the anti-Hermitian cluster operators $\sigma_{\rm int}$ and $\sigma_{\rm ext}$ are expressed in terms of parameters containing all active spin-orbital indices ($\sigma_{\rm int}$) and at least one inactive spin-orbital index ($\sigma_{\rm ext}$). More details can be found in Ref. \onlinecite{downfolding2020t}. A similar decomposition of the standard cluster operator is employed in active-space CC formulations, see Refs.\citenum{oliphant1991multireference,oliphant1992implementation,pnl93}.
Once certain conditions are met (associated with the convergence of infinite series as discussed in Ref.\onlinecite{downfolding2020t}) the DUCC Ansatz represents the exact ground-state wave function. The corresponding energy $E$ can be reproduced (approximated) once the operator $\sigma_{\rm ext}$ is know (or can be approximated) as an eigenvalue of the effective Hamiltonian $H^{\rm eff}$ given by the $\sigma_{\rm ext}$-dependent formula,
\begin{equation}
    H^{\rm eff} = (P+Q_{\rm int})
    e^{-\sigma_{\rm ext}} H
    e^{\sigma_{\rm ext}}
    (P+Q_{\rm int})\;.
    \label{eq2}
\end{equation}
Here $P$ and $Q_{\rm int}$ are the projection operators on the reference function $|\Phi\rangle$ (usually chosen as Hartree-Fock determinant) and all excited Slater determinants with respect to $|\Phi\rangle$ in the active space of interest ($P+Q_{\rm int}$ represents a projection operator on the active space).
Symbolically, we will represent the CC downfolding procedure, which downfolds Hilbert space $\cal{H}$ to active space ${\cal M}$ as
\begin{equation}
H \longrightarrow H^{\rm eff}({\cal H},H,{\cal M}) \;,
\label{eq3}
\end{equation}
where it is explicitly emphasized that the $H^{\rm eff}$ Hamiltonian downfolds correlation effects in the entire Hilbert space ${\cal H}$ defined by the Hamiltonian $H$.

In practical applications, an approximate form of the $H^{\rm eff}$ is used based on the finite-rank commutator expansion. In this Communication we will use the A(7) type approximation discussed in Ref. \onlinecite{doublec2022}.
Additionally, we approximate the $\sigma_{\rm ext}$ operator using the unitary coupled cluster formalism, which leverages cluster amplitudes obtained from the standard CCSD simulations, and we restrict the many-body form of $H^{\rm eff}$  to one- and two-body interactions.

\subsection{Quantum Flow}

The QFlow algorithm can be derived by analyzing the energy functional defined for the so-called {\em target active space} $\cal{M}_{\rm t}$ 
 (${\cal M}_{\rm t} \subset {\cal H}$) which is considered to be too large to be handled by available quantum resources. The energy functional is given by, 
\begin{equation}
E = \langle\Phi| e^{-\sigma_{\rm QF}} H e^{\sigma_{\rm QF}} |\Phi\rangle
\label{var1}
\end{equation}
where one assumes that the anti-Hermitian cluster operator $\sigma_{\rm QF}$ for the ground state of interest can be represented as a superposition of anti-Hermitian  cluster operators
$\sigma_{\rm int}(i)$ defined for 
various complete active spaces (CASs) $\lbrace {\cal M}_i\rbrace_{i=1}^{M}$, i.e.,
\begin{equation}
  \sigma_{\rm QF} \simeq 
  {\widetilde{\sum}}_{i=1}^M 
  \sigma_{\rm int}(i) \;,
  \label{var2}
\end{equation}
where ${\widetilde{\sum}}_{i=1}^M \sigma_{\rm int}(i)$  denotes a unique combination of the anti-Hermitian cluster operators associated with the individual active spaces included in the flow (see Refs. \cite{kowalski2023quantum,kowalski2025resource} for further  details). 
We will analyze the functional (\ref{var1}) from the perspective of the $i$-th active space. To this end, we represent the $\sigma_{\rm QF}$ operator using a rank-$N$ Trotter decomposition,
\begin{equation}
 \sigma_{\rm QF} =  \sigma_{\rm int}(i)+ \sigma_{\rm ext}(i) \simeq ( e^{\sigma_{\rm ext}(i)/N}
     e^{\sigma_{\rm int}(i)/N})^{N} \;.
 \label{var3}
\end{equation}
Using this expansion, we can re-cast the optimization problem (\ref{var1}) as a set of coupled minimization problems. For example, for the rank-1 ($N=1$) Trotter decomposition, one gets the following form of the Hermitian quantum flow,
\begin{equation}
\min_{\sigma_{\rm int}(i)} \langle\Phi| e^{-\sigma_{\rm int}(i)}H^{\rm eff}(i) e^{\sigma_{\rm int}(i)}|\Phi\rangle  \;,\; (i=1,\ldots,M) \label{var4} 
\end{equation}
where each problem is optimized with respect to amplitudes defining the $\sigma_{\rm int}(i)$ operator.
For rank-1 Trotter decomposition of $e^{\sigma}$/$e^{-\sigma}$ with respect to the partitioning (\ref{var3}), the 
effective Hamiltonians $H^{\rm eff}(i)$ are defined as
\begin{equation}
H^{\rm eff}(i)=(P+Q_{\rm int}(i)) e^{-\sigma_{\rm ext}(i)} H  e^{\sigma_{\rm ext}(i)} (P+Q_{\rm int}(i)) \;, \label{var5}
\end{equation}
where $H^{\rm eff}(i)$ is obtained by the downfolding procedure within the ${\cal M}_{\rm t}$ space, i.e., 
\begin{equation}
H \longrightarrow H^{\rm eff}(i)
= H^{\rm eff}({\cal M}_t,H,{\cal M}_i)
\label{var6}
\end{equation}
In previous QFlow simulations \cite{kowalski2023quantum,kowalski2025resource} we assumed that each $\sigma_{\rm int}(i)$ is represented in the form of standard unitary CC (UCC) cluster operator,  i.e.,
\begin{equation}
\sigma_{\rm int}(i) \simeq T_{\rm int}(i) -
T_{\rm int}(i)^{\dagger} \;.\label{var2a}
\end{equation}
The same convention will be employed in the present studies.

The flow equations can be solved using the importance ordering of the active spaces (defined with respect to the strength of encapsulated ground-state correlation effects) and optimization of the unique amplitudes, which, due to the possibility of overlapping active spaces, have not been optimized in preceding active-space problems. As discussed in Refs.\cite{kowalski2021dimensionality,bauman2022coupled,kowalski2023quantum}, the energy of QFlow is recorded for the main active space (or the active space no. 1 in the hierarchy of active spaces), which means that the remaining active-space problems can be viewed as "learning agents" to learn correlation effects outside of the main active space.
In this work, we also examine alternative formulations for defining the QFlow energy.
Since equations for each active space are expressed in terms of connected diagrams (which can be demonstrated using commutator expansion),  the QFlow energies are size-consistent.

The QFlow procedure can  also be a part of two-step downfolding  procedure defined as 
\begin{widetext}
\begin{eqnarray}
    H & \longrightarrow & H^{\rm eff}({\cal H},H,{\cal M}_{\rm t}) \;, \label{eq10} \\
    H^{\rm eff}({\cal H},H,{\cal M}_{\rm t}) 
    & \longrightarrow & 
    H^{\rm eff}({\cal M}_{\rm t},H^{\rm eff}({\cal H},H,{\cal M}_{\rm t}),{\cal M}_i) \;\; (i=1,\ldots,M)\;\;,
    \label{eq11}
\end{eqnarray}
\end{widetext}
where ${\cal M}_{\rm t}$-related QFlow algorithm instead of using bare Hamiltonian $H$ in ${\cal M}_{\rm t}$ now utilizes the effective Hamiltonian that correspond to the downfolding of correlation effect from entire Hilbert space ${\cal H}$ to the target space ${\cal M}_{\rm t}$.

\subsection{ADAPT-VQE Benchmark}

In order to benchmark the results of the QFlow procedure, we compare the energies to those obtained using standard ADAPT-VQE techniques \cite{grimsley2019adaptive,tang2021qubit,feniou2023overlap}. The ADAPT-VQE algorithm iteratively constructs the Ansatz,
\begin{equation}
    \ket{\psi^{\mathrm{ADAPT}}(\vec{\theta})} = \prod_{k = 0}^{N-1} \left( e^{\theta_{N-k}A_{N-k}} \right) \ket{\psi^{\mathrm{HF}}}
\end{equation}
over $N$ steps. 
At each step, all operators in a predefined pool are tested to select the operator $A_k$ that maximizes the gradient. The selected operator is added to the ansatz with a new variational parameter, $\theta_k$. The ADAPT step is completed by performing a VQE implementation to minimize of the cost function $\bra{\psi(\vec{\theta})}H\ket{\psi(\vec{\theta})}$ by optimizing over all parameters $\vec{\theta}$. Convergence is reached when the largest gradient has fallen below a target threshold. For our implementation, the operator pool consists of all singles, $A_p^q = a_{p}^\dagger a_{q} - a_{q}^\dagger a_{p}$, and doubles, $A_{rs}^{pq} = a_{p}^\dagger a_{r}^\dagger a_{q} a_{s} - a_{s}^\dagger a_{q}^\dagger a_{r} a_{p}$, excitation operators preserving spin and orbital symmetries.

\section{Implementation}
The matrix-based formulation of the QFlow algorithm begins with a choice of representation for the many-body Hilbert space. Within this framework, the algorithm is expressed using explicit matrix representations of all second-quantized operators entering the construction of the effective Hamiltonian. These matrix representations are built using the occupation number formalism, in which each $N$-electron Slater determinant basis state is uniquely specified by a binary string of length $M$,
\[
\lvert n \rangle = \lvert n_M\, n_{M-1} \cdots n_{i+1}\, n_i\, n_{i-1} \cdots n_1 \rangle
\tag{14}
\]
where each occupation number $n_i \in \{0,1\}$ indicates whether the $i$-th spin-orbital is occupied ($n_i =1$) or vacant ($n_i = 0$). The total number of spin-orbitals is $M$ = $2$n, with $n$ denoting the number of spatial orbitals. The fermionic creation and annihilation operators act on these basis states according to 
\begin{widetext}
\begin{align}
a_i^\dagger \lvert n_M \cdots n_{i+1}\, 0\, n_{i-1} \cdots n_1 \rangle 
&= (-1)^{\sum_{k=1}^{i-1} n_k} 
\lvert n_M \cdots n_{i+1}\, 1\, n_{i-1} \cdots n_1 \rangle 
\tag{15} \\
a_i \lvert n_M \cdots n_{i+1}\, 1\, n_{i-1} \cdots n_1 \rangle 
&= (-1)^{\sum_{k=1}^{i-1} n_k} 
\lvert n_M \cdots n_{i+1}\, 0\, n_{i-1} \cdots n_1 \rangle. 
\tag{16}
\end{align}
\end{widetext}
where the phase factor $(-1)^{\sum_{k=1}^{i-1} n_k}$ accounts for the fermionic anticommutation relations by counting the number of electrons occupying spin-orbitals with indices less than $i$. All other actions of $a_i^\dagger$ and $a_i$, attempting to create an electron in an already occupied orbital or annihilate an electron from a vacant one yield zero consistent with the Pauli exclusion principle. With the operator actions on the occupation number basis fully defined by equations (15) and (16), explicit matrix representations of all operators entering the QFlow algorithm can be constructed systematically over the full many-body Fock space. Specifically, the electronic Hamiltonian, the external cluster operator, the corresponding unitary transformations, the similarity-transformed Hamiltonian, and the active space effective Hamiltonian are each represented as matrices: 
\begin{align}
H &\rightarrow \mathbf{H}, \tag{17} \\
T_{\text{ext}} &\rightarrow \mathbf{T}_{\text{ext}}, \tag{18} \\
e^{\sigma_{\text{ext}}} 
&\simeq e^{T_{\text{ext}} - T^\dagger_{\text{ext}}} 
\rightarrow e^{\mathbf{T}_{\text{ext}} - \mathbf{T}^\dagger_{\text{ext}}}, \tag{19} \\
e^{-\sigma_{\text{ext}}} 
&\simeq e^{-(T_{\text{ext}} - T^\dagger_{\text{ext}})} 
\rightarrow {e}^{-(\mathbf{T}_{\text{ext}} - \mathbf{T}^\dagger_{\text{ext}})}, \tag{20} \\
\bar{H}_{\text{ext}} &\rightarrow \bar{\mathbf{H}}_{\text{ext}}, \tag{21} \\
H^{\text{eff}} &\rightarrow \mathbf{H}^{\text{eff}}. \tag{22}
\end{align}
The similarity-transformed Hamiltonian $\bar{H}_{\mathrm{ext}}$ is constructed explicitly in matrix form by applying the external similarity transformation to the Hamiltonian matrix. For each active space $M_i$, an effective Hamiltonian $H^{\mathrm{eff}}(i)$ is obtained by extracting the $i$-th active space sub-block of $\bar{H}_{\mathrm{ext}}(i)=e^{-\sigma_{\rm ext}(i)}He^{\sigma_{\rm ext}(i)}$.
This step constitutes the central operation 
The internal unitary cluster operator $\sigma_{\mathrm{int}}(i)$ is then used to variationally parameterize the active-space wave function, and optimization over its amplitudes yields the ground-state energy for that block, along with the amplitudes propagated forward in the flow.

For the  ADAPT-VQE validations, we have implemented a circuit simulation using the NWQ-VQE module of the NWQSim quantum simulation environment \cite{nwqsim-repo}. The internal VQE implementations within each ADAPT step were carried out with a derivative-based, limited-memory BFGS optimizer using a simultaneous perturbation stochastic approximation (SPSA) gradient method due to the large size of the parameter space. The gradient tolerance for convergence was set to $10^{-3}$.

\section{Results and Discussion}
To assess the performance of the QFlow algorithm, we carried out simulations on a set of representative chemical systems represented by various basis sets. 

\subsection{H8}

\begin{figure*}[t]
    \centering
    \includegraphics[width=0.8\linewidth]{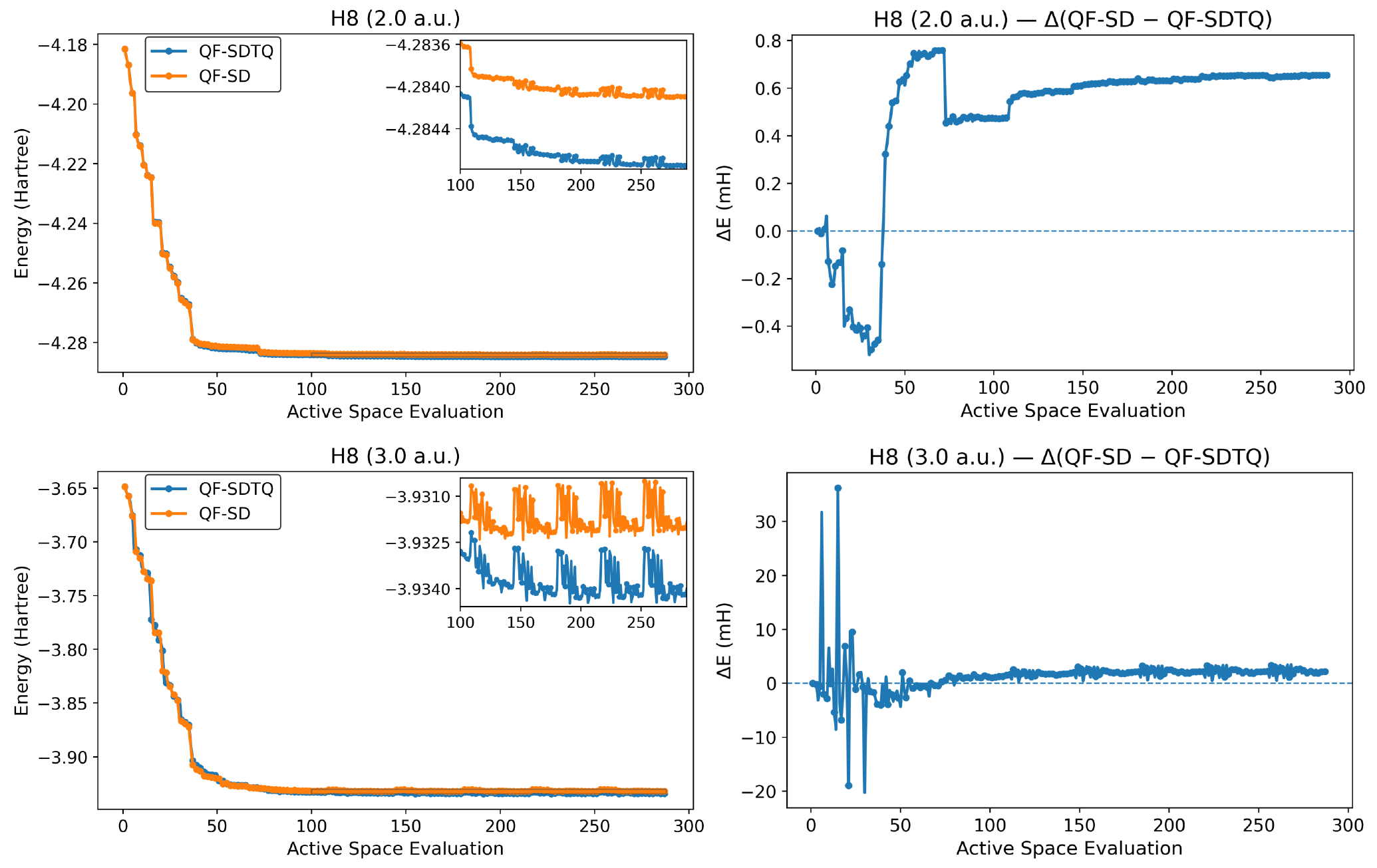}
    \caption{Active space energy profiles for the QFlow calculation of H8 with 2.0 and 3.0 a.u. separations using the STO-3G basis. Eight cycles, each comprising 36 active spaces, are shown. Figures on the left show the total energy (in Hartree) of each active space as the algorithm progresses, and the subgraphs show the energies after 100 active-space evaluations. The figures on the right show the energy difference (in milliHartree) between the QF-SD and QF-SDTQ methods for each active-space evaluation.}
    \label{fig:H8}
\end{figure*}

The first system investigated is H8, consisting of eight hydrogen atoms equidistantly separated along a line. For the H8 system, we investigated both the weakly 
correlated regime, where the hydrogen separation is 2.0 a.u., and the strongly correlated regime, where the separation is 3.0 a.u., using the STO-3G basis \cite{hehre1969self}. The QFlow algorithm casts the parent system of eight electrons in eight orbitals as
36 active spaces corresponding to different combinations of four electrons and four orbitals (4e,4o). This reduces the qubit requirement from 16 qubits in the parent problem 
to 8 qubits for each subspace problem. The (4e,4o) subspace is the smallest active space that one can use to cover
all single and double excitations, as well as a subset of triples and quadruples if they are included in the variational ansatz. It's 
important to recall that the qubit requirement of the parent problem is dictated by the choice of the basis set representing the system
which can correspond to hundreds or thousands of qubits, while the qubit requirement for QFlow calculations is based on the size of the largest
subspace, which can result in substantial savings compared to the requirements for the parent calculations, especially with larger basis sets. 

\begin{figure}[t]
    \centering
    \includegraphics[width=0.9\linewidth]{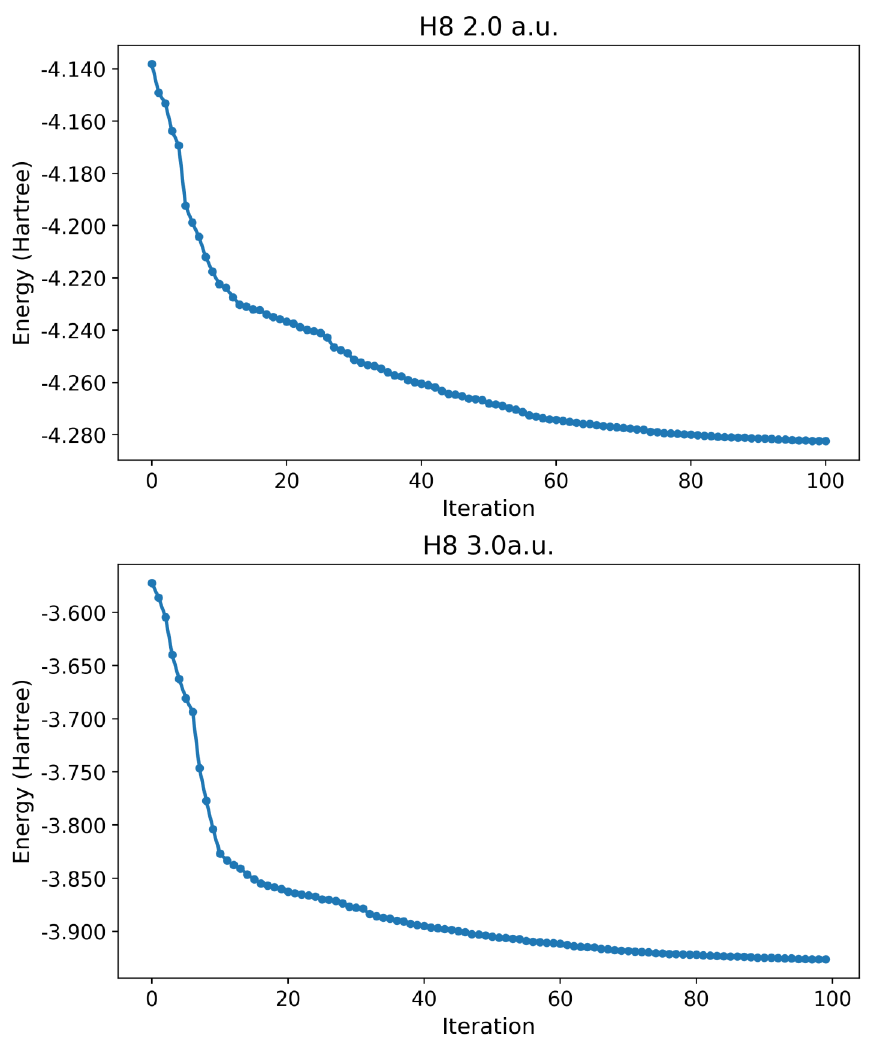}
    \caption{Energy convergence of the ADAPT-VQE calculations of H8 with 2.0 and 3.0 a.u. separations using the STO-3G basis. The gradient threshold for convergence is set to $10^{-3}$ with a central difference gradient step size of $10^{-4}$.}
    \label{fig:H8-VQE}
\end{figure}

QFlow results with the singles-and-doubles parameterization (QF-SD) and the higher-level analog with up to quadruple excitations (QF-SDTQ) for the (4e,4o) active space are shown in Figure \ref{fig:H8} and summarized in Table \ref{tab:H8energies}. The table also includes self-consistent-field (SCF) and exact diagonalization (ED) energies for comparison, along with final UCCSD values from ADAPT-VQE calculations shown in Figure \ref{fig:H8-VQE}. In the weakly correlated regime (2.0 a.u.), the full UCCSD ADAPT-VQE energy agrees well with the ED value, indicating that higher-order correlations play a minor role at this geometry. As a result, we see that each active-space evaluation for the QF-SD approximation closely follows the QF-SDTQ ansatz, with a difference of less than 1 mH for each active space. After the third cycle, the active space energies are well converged, and the remaining cycles show a nearly constant shift of $\sim$0.6 mH, with the QF-SD approach being the higher energy approximation. Whether the mean, median, highest-energy active space, lowest-energy active space, or any particular active space should be used as the representative energy for a given cycle remains unclear. Luckily, all of these statistics lie within 1 mH between the two QFlow approximations for the 2.0 a.u. separation. For the strongly correlated case (3.0 a.u.), the UCCSD energy is over 10 mH above ED, indicating the importance of higher-order correlations, which is why differences between the QF-SD and QF-SDTQ methods are more noticeable but still very systematic. The first few cycles show larger differences between the corresponding active-space energies, as seen in the last panel of Figure \ref{fig:H8}. By the end of the third cycle, the energies have stabilized and, as before, the QF-SD approach converges to higher energies than QF-SDTQ. The largest difference for any one active space is 3 mH, while the remaining energy statistics differ by $\sim$2 mH between the two ansatzes.

\begin{table}[ht]
\begin{center}
\caption{QFlow energy statistics\protect\footnotemark[1] and benchmark energies (Hartree) for the H8 system at
2.0 a.u. and 3.0 a.u. separation distances using the STO-3G basis.}
\label{tab:H8energies}
\begin{tabular}{r@{\hspace{12pt}} r@{\hspace{12pt}} c@{\hspace{12pt}} c}
\hline \hline
\multicolumn{1}{c}{Method}  &  & 2.0 a.u. & 3.0 a.u. \\ 
\hline
SCF     &          & -4.1382 & -3.5723 \\
UCCSD   &          & -4.2841 & -3.9341 \\
QF-SD   & Primary\protect\footnotemark[2]  & -4.2840 & -3.9305 \\
        & Mean     & -4.2841 & -3.9316 \\
        & Median   & -4.2841 & -3.9319 \\
        & Std. Dev.& 2.98E-05 & 5.98E-04 \\
        & Min      & -4.2841 & -3.9323 \\
        & Max      & -4.2840 & -3.9305 \\
QF-SDTQ & Primary\protect\footnotemark[2]  & -4.2847 & -3.9327 \\
        & Mean     & -4.2847 & -3.9338 \\
        & Median   & -4.2847 & -3.9340 \\
        & Std. Dev.& 3.23E-05 & 5.85E-04 \\
        & Min      & -4.2848 & -3.9345 \\
        & Max      & -4.2847 & -3.9327 \\
ED      &          & -4.2860 & -3.9447 \\
\hline \hline
\end{tabular}
\vspace{-1 em}
\footnotetext[1]{
\setlength{\baselineskip}{1em}
Energy statistics are reported for the last cycle of active spaces. A total of eight cycles were run.}
\footnotetext[2]{
\setlength{\baselineskip}{1em}
Primary refers to the active space consisting of the two highest energy occupied orbitals and two lowest energy unoccupied orbitals.}
\end{center}
\end{table}

 \subsection{H$_2$O} \label{H2O_sub_sec}
The next system is H$_2$O, where we investigated the equilibrium geometry 
($\mathrm{r}(\mathrm{O\!-\!H}) = 0.9578\,\text{\AA} \quad \text{and} \quad \angle(\mathrm{H\!-\!O\!-\!H}) = 104.4775^\circ$) 
as well as geometries where
one of the bonds was stretched to 1.5 and 2.0 times the equilibrium bond length. We employed the cc-pVTZ basis, but reduced the dimensionality of the problem using two forms of the 
Hamiltonians -- bare, in which we truncated the restricted Hartree--Fock Hamiltonian to an (8e,8o) active space and DUCC effective Hamiltonian for the same-size active space, which is based on the A(7) type approximation discussed in Ref. \onlinecite{doublec2022}. The QFlow algorithm utilized 36 combinations (4e,4o) active spaces.

\begin{figure}[t]
    \centering
    \includegraphics[width=0.9\linewidth]{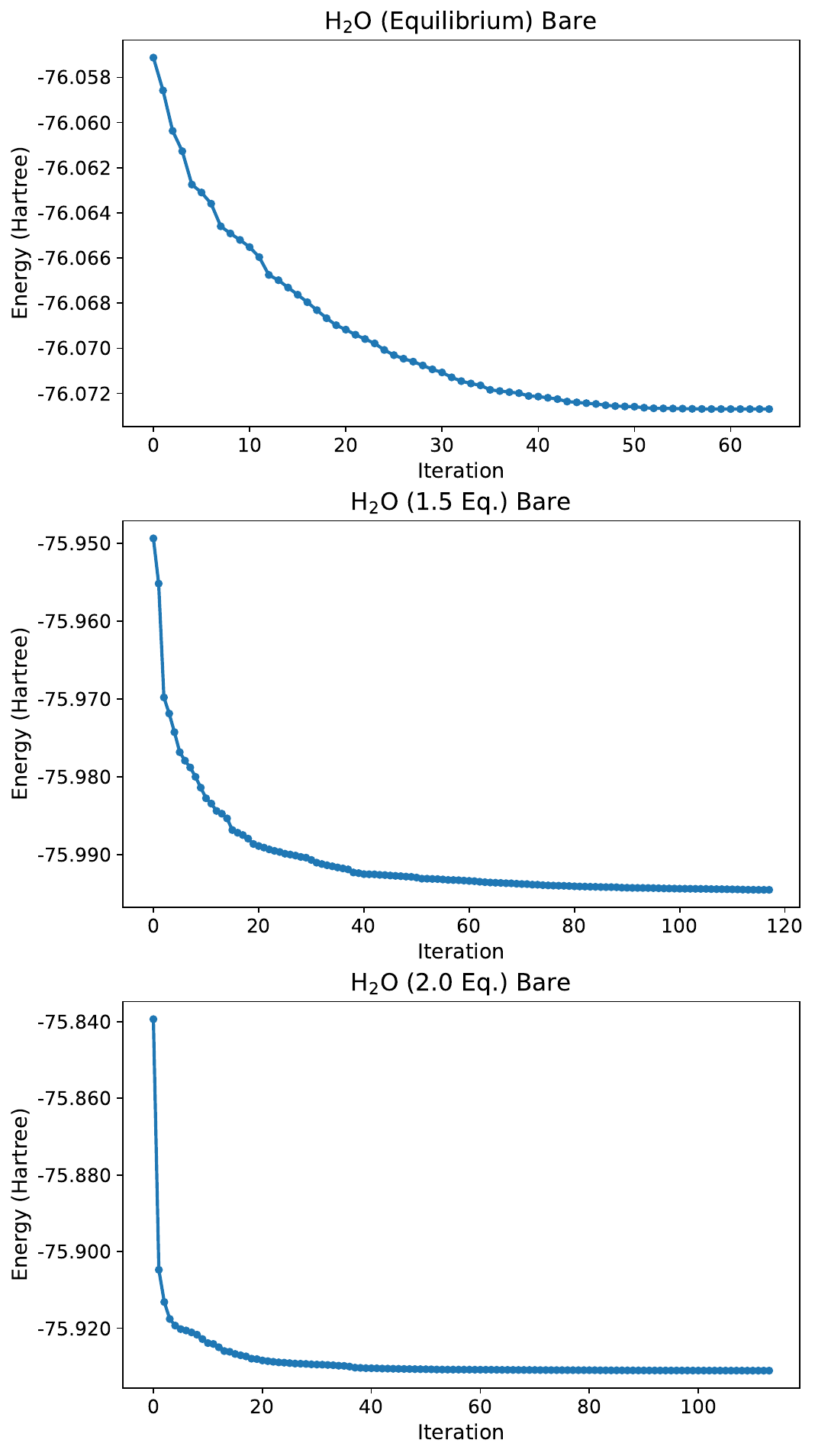}
    \caption{Energy convergence of the ADAPT-VQE calculations of H$_2$O at equilibrium and single bond stretched geometries using the bare Hamiltonian. The gradient threshold for convergence is set to $10^{-3}$ with a central difference gradient step size of $10^{-4}$.}
    \label{fig:H2O-Bare-VQE}
\end{figure}

\begin{figure}[t]
    \centering
    \includegraphics[width=0.9\linewidth]{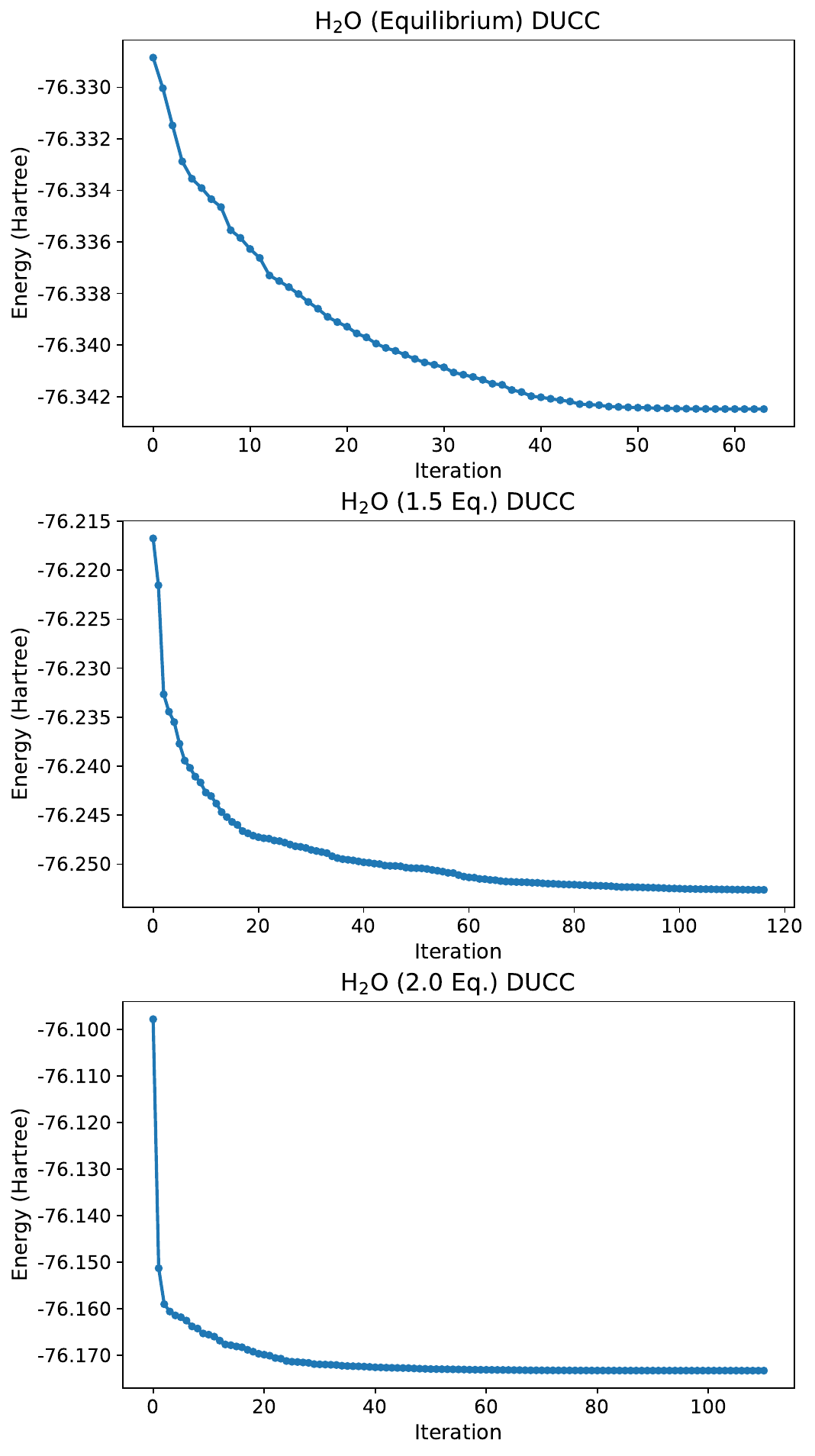}
    \caption{Energy convergence of the VQE calculations of H$_2$O at equilibrium and single bond stretched geometries using the DUCC downfolded Hamiltonian. The gradient threshold for convergence is set to $10^{-3}$ with a central difference gradient step size of $10^{-4}$.}
    \label{fig:H2O-DUCC-VQE}
\end{figure}

At equilibrium, the electron correlation in H$_2$O is largely dynamical, and interactions among all orbitals play a role in describing the wavefunction. This is reflected in the ADAPT-VQE calculation, where there is a relatively slow but steady convergence in energy as the calculation iterates, as seen in Figures \ref{fig:H2O-Bare-VQE} and \ref{fig:H2O-DUCC-VQE}. As the O--H bond is stretched, a few wavefunction parameters dominate as we transition to a more strongly correlated regime. This correlation is characterized by large single- and double-excitation amplitudes, and ADAPT-VQE picks it up early, as we observe more pronounced relative and total energy changes at the beginning of the calculations. Higher-order correlations contribute much less to the system, as the final UCCSD energies differ from the ED values by 0.5 mH or less for all geometries. 

\begin{figure*}[ht]
    \centering
    \includegraphics[width=\linewidth]{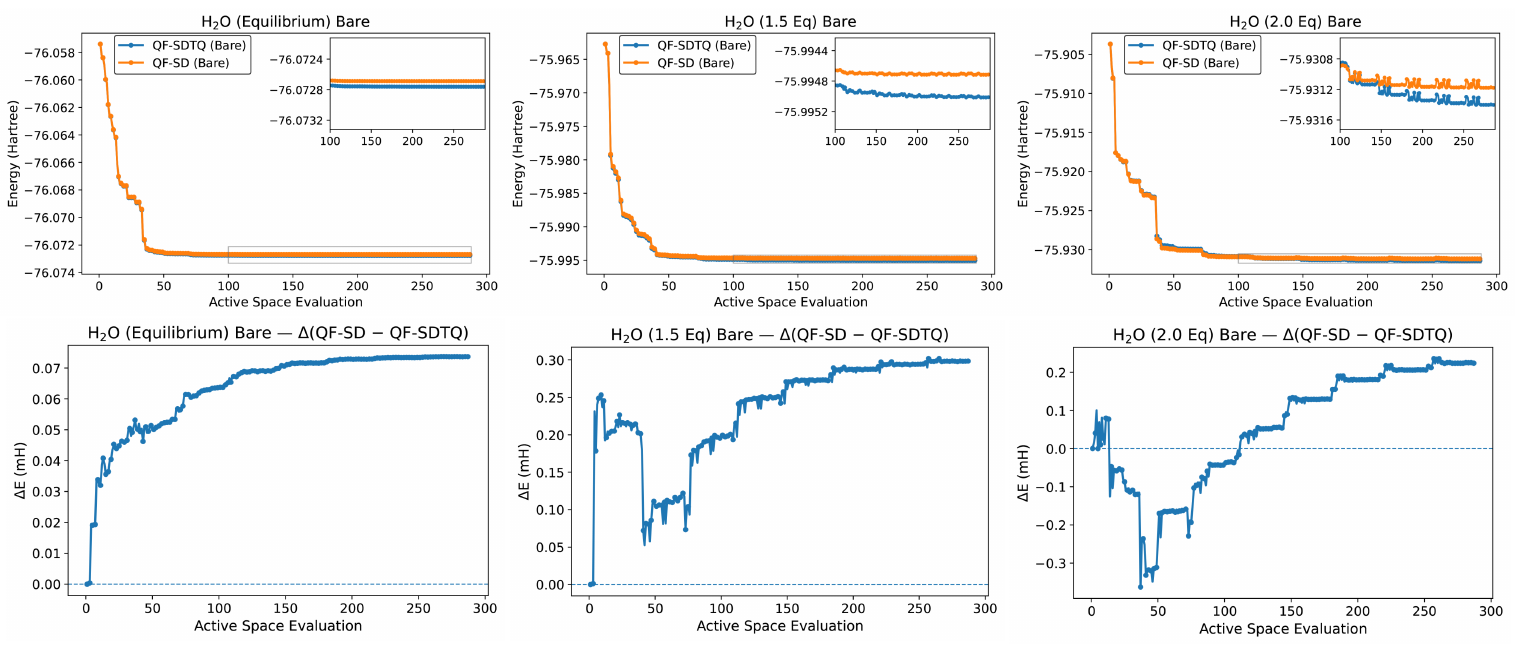}
    \caption{Active space energy profiles for the QFlow calculation of H$_2$O at equilibrium and single bond stretched geometries, obtained using the truncated bare Hamiltonians. Details of the basis and selected orbitals are found in Section~\ref{H2O_sub_sec}. Eight cycles, each comprising 36 active spaces, are shown. Figures on the top row show the total energy (in Hartree) of each active space as the algorithm progresses, and the subgraphs show the energies after 100 active-space evaluations. The figures on the bottom row show the energy difference (in milliHartree) between the QF-SD and QF-SDTQ methods for each active-space evaluation.}
    \label{fig:H2O-Bare}
\end{figure*}

The QFlow results with the bare Hamiltonians are shown in Table \ref{tab:H2O_Bare_energies} and Figure \ref{fig:H2O-Bare}. Since the correlation is largely characterized by single- and double-excitation contributions, the QF-SD approach is expected to perform well, and this is indeed the case when compared to the QF-SDTQ method. The difference between the two methods for the three geometries is 0.3 mH or less. This agreement allows us to focus on how the two approaches cycle through the active spaces. The first observation is that the difference between QF-SD and QF-SDTQ converges quickly at equilibrium, but takes longer as the bond length is stretched and correlation becomes stronger. Upon close inspection, we see that the QF-SD approximation, with the smaller parameter space, actually converges faster and is well converged by the 5th or 6th cycle. It is the QF-SDTQ approach that is undergoing changes in the later cycles. The second observation is that within a cycle, we have two groupings of energies which become more separate the further the O--H bond is stretched. The amplitudes describing the bond breaking become very large at stretched geometries, and for some active spaces, they are part of the internal subset $\sigma_{int}$, while for other active spaces, they belong to the external subset $\sigma_{ext}$. Whether these amplitudes are solved for as eigenvectors of the effective Hamiltonians (belonging to the internal subset) or used to construct the effective Hamiltonian (belonging to the external subset) gives rise to the two groupings of active-space energies. In the limit where exact effective Hamiltonians can be formed, all active spaces would have the same energy. This observation is further important when discussing the results for the DUCC Hamiltonians.

\begin{table}[H]
\begin{center}
\begin{threeparttable}
\caption{QFlow energy statistics$^{\rm a}$ and benchmark energies (Hartree) for H$_2$O with the bare Hamiltonian of the cc-pVTZ basis truncated to eight lowest-energy orbitals ($\dim({\cal M}_{\rm t})=8$).}
\label{tab:H2O_Bare_energies}
\begin{tabular}{r@{\hspace{12pt}} r@{\hspace{12pt}} c@{\hspace{12pt}} c@{\hspace{12pt}} c}
\hline \hline
\multicolumn{1}{c}{Method}  &  & Eq. & 1.5*Eq. & 2.0*Eq. \\
\hline
SCF     &                      & -76.0571                & -75.9494                    & -75.8393 \\
UCCSD   &                      & -76.0727                & -75.9945                    & -75.9311 \\
QF-SD   & Primary$^{\rm b}$ & -76.0727                & -75.9947                    & -75.9311 \\
        & Mean                 & -76.0727                & -75.9947                    & -75.9311 \\
        & Std. Dev.            & 1.60E-07                & 7.65E-06                    & 3.51E-05 \\
QF-SDTQ & Primary$^{\rm b}$ & -76.0728                & -75.9950                    & -75.9313 \\
        & Mean                 & -76.0728                & -75.9950                    & -75.9314 \\
        & Std. Dev.            & 1.68E-07                & 7.56E-06                    & 3.65E-05 \\
ED      &                      & -76.0729                & -75.9952                    & -75.9315 \\
\hline \hline
\end{tabular}
\begin{tablenotes}
\footnotesize
\item[$^{\rm a}$] Energy statistics are reported for the last cycle of active spaces. 
A total of eight cycles were run.
\item[$^{\rm b}$] Primary refers to the active space consisting of the two highest 
energy occupied orbitals and two lowest energy unoccupied orbitals.
\end{tablenotes}
\end{threeparttable}
\end{center}
\end{table}

\begin{figure*}[t]
    \centering
    \includegraphics[width=\linewidth]{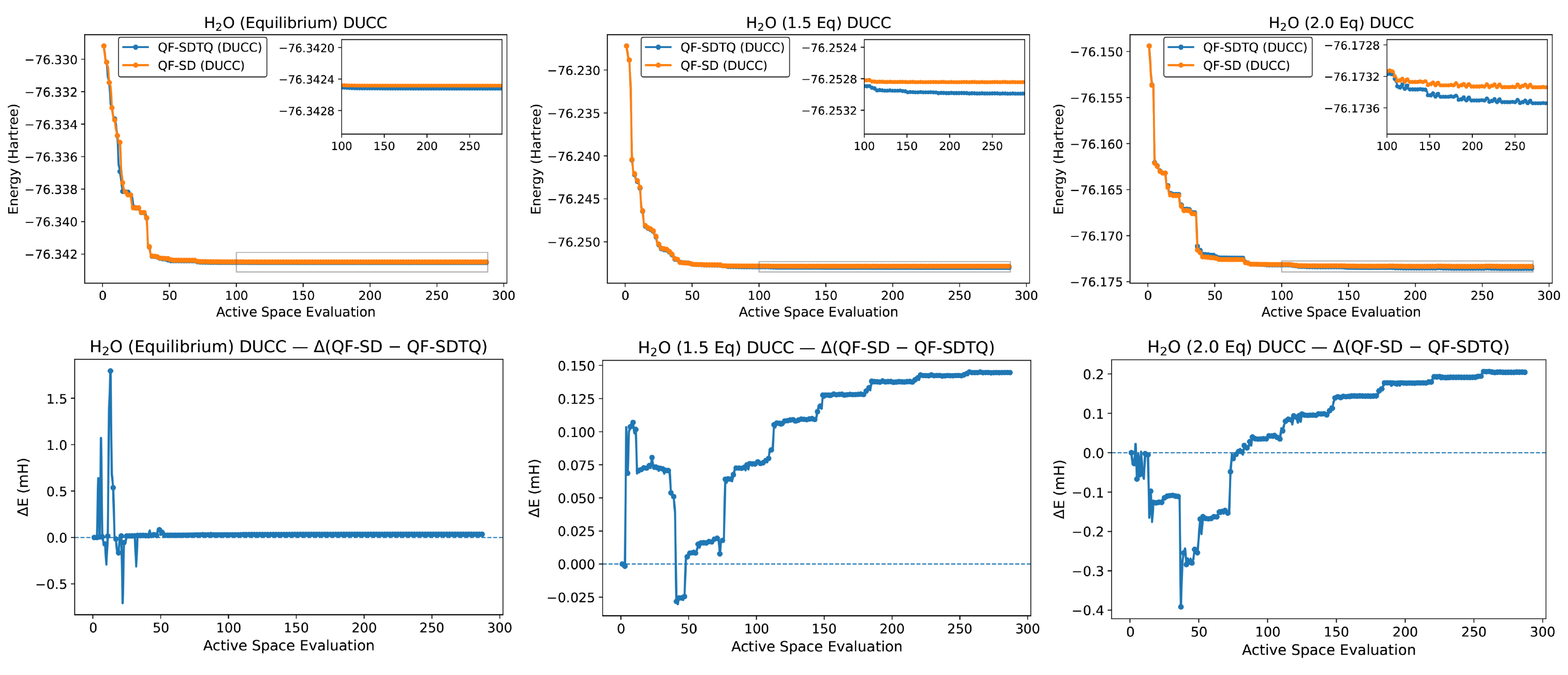}
    \caption{Active space energy profiles for the QFlow calculation of H$_2$O at equilibrium and single bond stretched geometries, obtained using the DUCC downfolded/effective Hamiltonians. Details of the basis and selected orbitals are found in Section~\ref{H2O_sub_sec}. Eight cycles, each comprising 36 active spaces, are shown. Figures on the top row show the total energy (in Hartree) of each active space as the algorithm progresses, and the subgraphs show the energies after 100 active-space evaluations. The figures on the bottom row show the energy difference (in milliHartree) between the QF-SD and QF-SDTQ methods for each active-space evaluation.}
    \label{fig:H2O-DUCC}
\end{figure*}

The QFlow results for H$_2$O with the DUCC Hamiltonians are shown in Table \ref{tab:H2O_DUCC_energies} and Figure \ref{fig:H2O-DUCC}. Once again, we observe excellent agreement between the QF-SD and QF-SDTQ approaches, with differences of 0.2 mH or less for the final cycles. In addition, there is a unique feature of DUCC effective Hamiltonians that has been studied in previous work\cite{DUCC-DMRG}, which is that the largely dynamical correlation brought by the effective Hamiltonian increases entanglement between orbitals and enhances the single-reference character in the active space. The wave function character shifts from a few dominant excitations to a more dispersed characterization, with excitations across more orbitals playing a larger role than in the bare Hamiltonian for the same active space. For the ADAPT-VQE calculations, this behavior yields an energy profile with a larger radius of curvature, although it has little effect on the total number of iterations required. For the QFlow calculations, we observe very similar patterns of convergence for all cycles between the DUCC and bare Hamiltonians, but the DUCC Hamiltonians consistently have a smaller deviation among active spaces within a cycle because the effective Hamiltonians in the QFlow algorithm are more evenly treated due to the dispersed excitation character.

\begin{table}[ht]
\begin{center}
\caption{QFlow energy statistics\protect\footnotemark[1] and benchmark energies (Hartree) for the DUCC Hamiltonian of H$_2$O where the cc-pVTZ basis is downfolded to the eight lowest-energy orbitals ($\dim({\cal M}_{\rm t})=8$).}
\label{tab:H2O_DUCC_energies}
\begin{tabular}{r@{\hspace{12pt}} r@{\hspace{12pt}} c@{\hspace{12pt}} c@{\hspace{12pt}} c}
\hline \hline
\multicolumn{1}{c}{Method}  &  & Eq. & 1.5*Eq. & 2.0*Eq. \\ 
\hline
SCF     &                      & -76.3289                & -76.2168                    & -76.0979 \\
UCCSD   &                      & -76.3425                & -76.2526                    & -76.1733 \\
QF-SD   & Primary\protect\footnotemark[2] & -76.3425                & -76.2528                    & -76.1733 \\
        & Mean                 & -76.3425                & -76.2528                    & -76.1733 \\
        & Std. Dev.            & 1.75E-08                & 1.03E-06                    & 1.18E-05 \\
QF-SDTQ & Primary\protect\footnotemark[2] & -76.3425                & -76.2530                    & -76.1735 \\
        & Mean                 & -76.3425                & -76.2530                    & -76.1735 \\
        & Std. Dev.            & 2.95E-08                & 1.24E-06                    & 1.39E-05 \\
ED      &                      & -76.3426                & -76.2531                    & -76.1736 \\
\hline \hline
\end{tabular}
\vspace{-1 em}
\footnotetext[1]{
\setlength{\baselineskip}{1em}
Energy statistics are reported for the last cycle of active spaces. A total of eight cycles were run.}
\footnotetext[2]{
\setlength{\baselineskip}{1em}
Primary refers to the active space consisting of the two highest energy occupied orbitals and two lowest energy unoccupied orbitals.}
\end{center}
\end{table}

\subsection{C$_2$ and SiC}

To investigate the behavior of the QFlow algorithm with periodic Hamiltonians, we considered the C$_2$ and SiC systems. The single-particle basis employed is obtained from the correlation-optimized virtual orbitals (COVOs)~\cite{ bylaska2021quantum, song2023periodic, inprepcovos} procedure, a new algorithm for generating a virtual space in which the orbitals are generated by minimizing small pairwise CI Hamiltonians. All the COVOs simulations in this study were performed using the pseudopotential plane-wave NWPW module~\cite{bylaska2011large}  implemented in the NWChem software package\cite{apra2020nwchem}. Restricted calculations for C$_2$ and SiC were carried out using Hartree-Fock exact exchange at bond lengths of 1.54~\AA\ and 1.89~\AA, respectively. The plane-wave calculations used a simple cubic box with L=15~\AA, and the electronic wavefunctions were expanded using a plane-wave basis under periodic boundary conditions with a wavefunction cutoff energy of 100 Ry and a density cutoff energy of 200 Ry. The valence electron interactions with the atomic core are approximated with generalized norm-conserving Hamann pseudopotentials ~\cite{hamann1989generalized, hamann1979norm} modified to the separable form suggested by Kleinman and Bylander ~\cite{kleinman1982efficacious, bylander1984outer}. The pseudopotentials used in this study were constructed using the following core radii: C: r$_{cs}$=0.800 a.u., r$_{cp}$=0.850 a.u., and r$_{cd}$=0.850 a.u.; Si: r$_{cs}$=1.059 a.u., r$_{cp}$=1.286 a.u., and r$_{cd}$=1.286 a.u. The HeteroFAM Web API (\url{https://heterofam.pnnl.gov/api/}), as a follow-up to the web application EMSL Arrows~\cite{bylaska2021building}, was used to set up and perform the plane-wave calculations.

\begin{figure*}[t]
    \centering
    \includegraphics[width=0.8\linewidth]{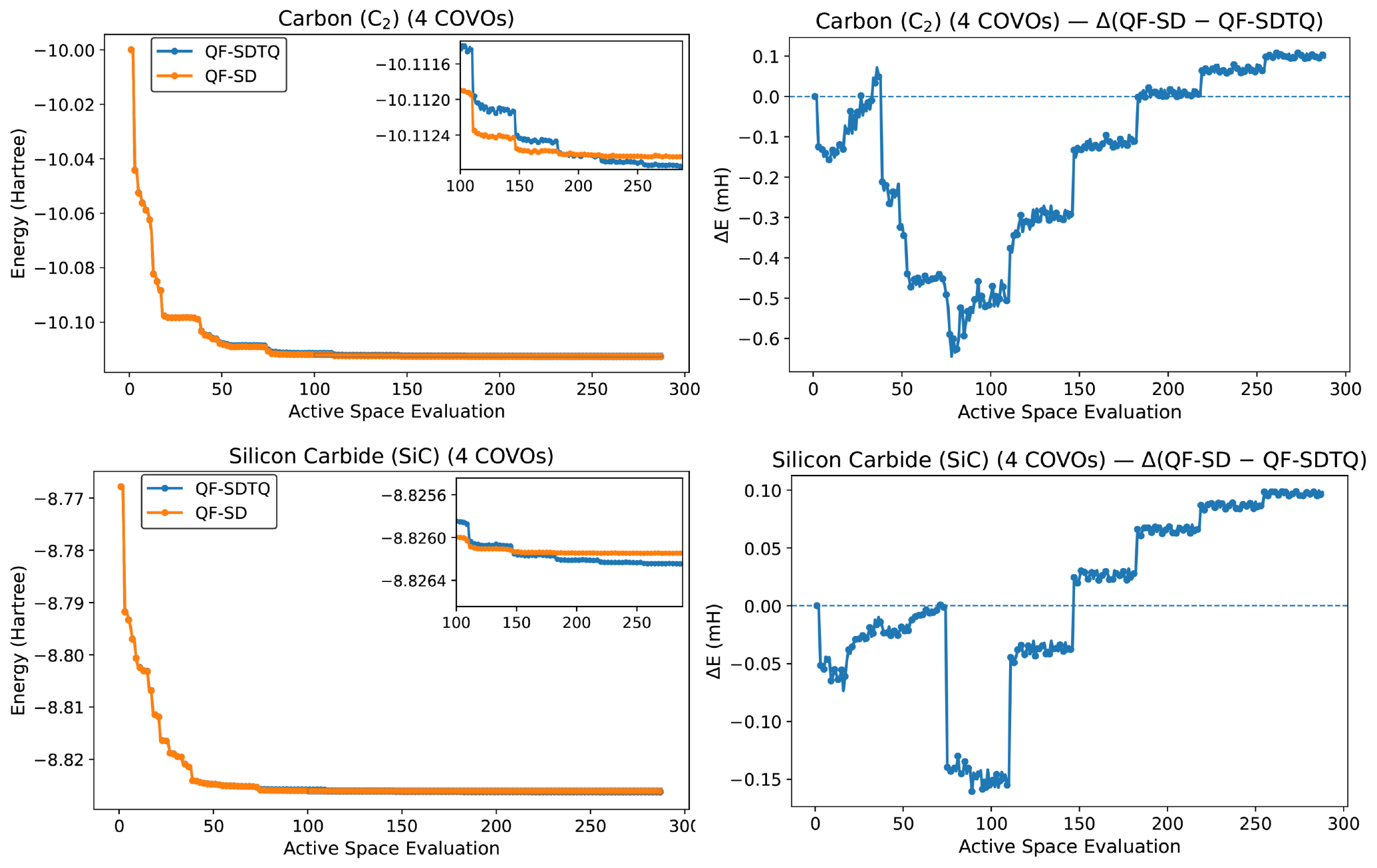}
    \caption{Active space energy profiles for the QFlow calculation of carbon (C$_2$) and silicon carbide (SiC) dimer using periodic COVOs-based Hamiltonians with (4e,4o) active spaces. Eight cycles, each comprising of 36 active spaces, are shown. Figures on the left show the total energy (in Hartree) of each active space as the algorithm progresses, and the subgraphs show the energies after 100 active-space evaluations. The figures on the right show the energy difference (in milliHartree) between the QF-SD and QF-SDTQ methods for each active space evaluation.}
    \label{fig:C2-SiC}
\end{figure*}

The QFlow calculations for C$_2$ and SiC, shown in Figure~\ref{fig:C2-SiC} and Table~\ref{tab:C2_SiC_energies}, recover 113 and 58 mH of correlation, respectively, which is close in magnitude to the H8 and H$_2$O stretched bond length calculations. We observed a similar pattern to those calculations, where the QF-SD ansatz has a faster descent followed by earlier convergence (by the 4th or 5th cycle) in the energy profile compared to the QF-SDTQ approach. The QF-SDTQ approach requires more cycles to converge, and we don't observe it falling below QF-SD until as late as the 6th cycle. Both approximations recover that same amount of correlation, only differing by 0.1 mH in the end, once again suggesting that the QFlow calculations with the singles and doubles ansatz can effectively recover the total electron correlation, including those for periodic Hamiltonians.
 
\begin{table}[ht]
\begin{center}
\caption{QFlow energy statistics\protect\footnotemark[1] and benchmark energies (Hartree) for C$_2$ and SiC with the periodic COVOs-based Hamiltonian.}
\label{tab:C2_SiC_energies}
\begin{tabular}{r@{\hspace{12pt}} r@{\hspace{12pt}} c@{\hspace{12pt}}  c}
\hline \hline
\multicolumn{1}{c}{Method}  &  & C$_2$ & SiC \\ 
\hline
QF-SD   & Primary\protect\footnotemark[2] & -10.1126               & -8.8261                 \\
        & Mean                 & -10.1126               & -8.8261                 \\
        & Std. Dev.            & 4.36E-06               & 8.92E-07                \\
QF-SDTQ & Primary\protect\footnotemark[2] & -10.1127               & -8.8262                 \\
        & Mean                 & -10.1127               & -8.8262                 \\
        & Std. Dev.            & 1.02E-05               & 2.51E-06                \\
\hline \hline
\end{tabular}
\vspace{-1 em}
\footnotetext[1]{
\setlength{\baselineskip}{1em}
Energy statistics are reported for the last cycle of active spaces. A total of eight cycles were run.}
\footnotetext[2]{
\setlength{\baselineskip}{1em}
Primary refers to the active space consisting of the two highest energy occupied orbitals and two lowest energy unoccupied orbitals.}
\end{center}
\end{table}

\section{Outlook}

While the capabilities of quantum devices continue to improve and expand the number of logical qubits available, it remains imperative to choose how to allocate those resources for best possible accuracy. Techniques such as CC downfolding allow for scaling to larger system sizes while reducing accuracy loss by incorporating information about the full many-body structure into the construction of the effective Hamiltonian within the active space. Even still, the size of the active space may be beyond the available qubit resources. In this case, the QFlow procedure provides a solution through the decomposition of the active space into CASs and recasting the minimization of the energy functional as a set of coupled minimization problems.       

Here we have demonstrated that the QFlow procedure with a low-rank excitation ansatz can effectively recover correlation energies for both chemical and materials-relevant periodic Hamiltonians, with substantially fewer qubits than required for VQE-UCC solvers covering the same excitation manifold. Furthermore, we highlight that within the QFlow framework, the singles and doubles ansatz is sufficient to capture a substantial portion of SDTQ-level correlation while using fewer variational parameters, further reducing computational costs. In addition, these results raise the possibility of an adaptive algorithm that leverages the low-cost and quick convergence of the singles and doubles excitations to capture the majority of the correlation, followed by selectively incorporating QF-SDTQ ansatz to refine the solution in regions where higher-order correlation is non-negligible.

To scale the application of the QFlow algorithm to the 10$^6$ parameter regime and beyond, we must transition from the matrix implementation to the many-body formulation, which introduces practical approximations. This work establishes that QFlow with a singles and doubles ansatz is a viable and foundational methodology for unlocking massively scaled quantum-parameter-space calculations. Furthermore, the modularity of the QFlow framework, grounded in the scalable many-body hierarchy of subsystem algebras, positions it as a practical and robust approach to high-accuracy quantum chemistry on emerging quantum computing architectures.                


\section{Acknowledgments}
This work was supported by the ``Embedding QC into Many-body Frameworks for Strongly Correlated Molecular and Materials Systems''  project, which is funded by the U.S. Department of Energy, Office of Science, Office of Basic Energy Sciences, the Division of Chemical Sciences, Geosciences, and Biosciences (under FWP 72689) and by Quantum Science Center (QSC), a National Quantum Information Science Research Center of the U.S. Department of Energy (under FWP  76213).
This work was partially supported by the U.S. Department of Energy, Office of Science, Basic Energy Sciences, Division of Materials Sciences and Engineering, Theoretical Condensed Matter Physics Program, FWP 83557.
KK and NPB also acknowledge the support from the 
Quantum Algorithms and Architecture for Domain Science Initiative (QuAADS), under the Laboratory Directed Research and Development (LDRD) Program at Pacific Northwest National Laboratory (PNNL). The COVOs calculations used resources of the National Energy Research Scientific Computing Center (NERSC), a User Facility supported by the Office of Science of the U.S. DOE under Contract No. DE-AC02-05CH11231.
This work used resources from the Pacific Northwest National Laboratory.
PNNL is operated by Battelle for the U.S. Department of Energy under Contract DE-AC05-76RL01830.

\section{Competing interests}
All authors declare no financial or non-financial competing interests. 


\bibliography{references}

\end{document}